\documentclass[aps,prl,reprint,groupedaddress]{revtex4-1}
\usepackage{amsmath, amsthm, amstext, amsfonts, hyperref, graphicx, braket}

\theoremstyle{definition}
\newtheorem*{defn}{Definition}

\begin{document}

\title{ 
The whole is greater than the sum of the parts:  \\
on the possibility of purely statistical interpretations of quantum theory}


\author{
Joseph Emerson, Dmitry Serbin, Chris Sutherland, Victor Veitch}
\affiliation{Institute for Quantum Computing, 200 University Ave West, Waterloo, ON, Canada}


\date{\today}

\begin{abstract}
The Pusey-Barrett-Rudolph theorem (PBR) claims to rule out the possibility of a purely statistical interpretation of the quantum state under an  assumption of how to represent independent operations in any hidden variable model.     
We show that PBR's assumption of independence encodes an assumption of local causality, which is already known to conflict with the predictions of quantum theory via Bell-type inequalities. 
We devise a weaker formulation of independence within a general hidden variable model that is empirically indistinguishable from the PBR assumption in situations where certain hidden variables are inaccessible.
Under this weaker principle we are able to construct an explicit hidden variable model that is purely statistical and also  reproduces the quantum predictions. Our results suggest that the assumption of a  purely statistical interpretation is actually an innocent bystander in the PBR argument, rather than the driving force behind their contradiction. 
\end{abstract}


\maketitle

\begin{quote}
``What is proved by the impossibility proofs is lack of imagination.''\cite{Bell82} 
\newline \em -- John S. Bell
\end{quote}

Perhaps the most long-standing open question in the foundations of quantum mechanics  is whether pure quantum states can be understood as states of incomplete knowledge about some underlying physical reality. That such an interpretation was possible, or even necessary, was central to Einstein's view of quantum theory: indeed, the EPR argument aimed to demonstrate that quantum states must be considered incomplete because of an  incompatibility between an assumption of locality  and the assumption that quantum states are complete a description of the physical states \cite{einstein1935can}.  Bell resolved that tension in the way that ``Einstein would have liked least" \cite{Bell66} by demonstrating that the assumption of hidden variables (and hence incompleteness) was also incompatible with an assumption of locality \cite{Bell64}.  In this way, EPR's idea that the properties of entangled states could be applied to disprove a particular type of interpretation was instead generalized by Bell into an insight about a violation of locality in a broad spectrum  of possible interpretations.

Recently, Pusey, Barrett and Rudolph (PBR) have devised a new impossibility proof showing that the predictions of quantum theory are incompatible with the following two assumptions \cite{pusey2012reality}: (1) quantum states are \emph{purely} statistical, in a sense we will make precise, and (2) a particular independence assumption about how independent quantum preparations must be represented in the hidden variables framework. The PBR argument considers a situation where each subsystem of a composite system is prepared independently, giving a product state preparation, after which the subsystems are brought together and then subjected to an entangling measurement. From the resulting conflict with the quantum predictions they \emph{infer}  that it is the assumption that quantum states are \emph{purely} statistical which is at fault and hence should be eliminated as a possibility. In subsequent work this inference is characterized as an \emph{implication} of their result \cite{Nigg2012,Barrett2013}. 
 Because the PBR assumption of preparation independence involves an assumption of locality, it is this assumption that should be regarded with utmost suspicion;  readers that consider locality assumptions natural in the hidden variable framework should consult this obscure paper: Ref.~\cite{Bell64}. Indeed the first contribution of this paper is to show that the PBR argument does not imply the impossibility of a purely statistical interpretation of quantum states, but instead demonstrates the impossibility of \emph{local causality} within these interpretations. There is, however, a positive upside to the PBR argument, or rather our analysis of it: we obtain novel  insight into the structure of hidden variable models that are consistent with the predictions of quantum theory. \\
\indent It is important to recognize that the PBR scenario is conceptually dual to the EPR-Bell scenario \cite{Bell64}: in the EPR-Bell set-up the states are entangled  and the measurements are independent, whereas in the PBR set-up the states are independent and measurements are  entangled.   Consequently it is natural to conjecture that the same  kind of ``locality" assumption can account for the conflict with quantum predictions in either scenario.   In the derivation of  Bell's theorem, it is widely accepted that the assumption of ``local causality" plays a critical role and, if one accepts the hidden variables framework, then it is this assumption that is violated by quantum mechanics. Below we first show that the PBR assumption of ``preparation independence" implicitly encodes the  assumption of local causality.  We then propose a weaker notion of independence that is appropriate in the framework of  hidden variables models: this notion, termed ``local independence'', is consistent with the no-signalling principle, and remarkably, is empirically indistinguishable from preparation independence when certain hidden variables are assumed to be inaccessible under local measurements.  We then construct an explicit hidden variable model for the PBR scenario to show that it is indeed possible to have a  purely statistical hidden variable interpretation of quantum states that reproduces the quantum predictions. 

The considerations of EPR and Bell, originally concerned with identifying how entanglement imposes surprising  constraints on possible interpretations of quantum theory, eventually inspired others to recognize the practical relevance of entanglement as a novel physical resource within the quantum formalism, for example, for achieving the advantages of quantum information. Similarly we hope that our analysis of the PBR ``no go" theorem, in addition to clarifying the foundational question of which interpretations of quantum theory are even possible, will also inspire further insight into the unique resources and practical capabilities of quantum theory.

\textit{Ontological models and hidden variables.}---  In Bell's ontological model framework the possible physical properties of a system  are identified with a space $\Lambda$. A preparation $\ket{\psi}$ is associated with a measure (probability density) $\mu_{\psi}(\lambda)$ over $\Lambda$ and each realization of the preparation results in a physical state $\lambda\in\Lambda$ sampled from $\mu_{\psi}(\lambda)$.  A measurement operator  $E_k$, e.g. drawn from a POVM $E=\{E_k\}$, is associated with a \textit{response function} $\xi_k(\lambda)$ which specifies the conditional probability of obtaining outcome $k$ given physical state $\lambda$. The conditional probabilities  $\xi_{k}(\lambda)$ and probabilities $\mu_{\psi}(\lambda)$ are subject to the following two conditions: (i) For all ontic states, some outcome always occurs $\forall\lambda\in\Lambda,\ \sum_{k}\xi_{k}(\lambda)=1$, and (ii) the Born rule statistics must be reproduced via the law of total probability $\text{Pr}(k|\ket{\psi})=\int_{\Lambda}\xi_{k}(\lambda)\mu_{\psi}(\lambda)\text{d}\lambda$.

Thus we associate an ontological model for quantum mechanics with the triplet ($\Lambda,\{\mu\},\{\xi\}$). This framework is very general. For example, it can accommodate the orthodox interpretation of von Neumann and Dirac in which quantum states are assumed to give a complete specification of the physical reality - that is, the denial of hidden variables: this interpretation is realized if we identify $\Lambda$ with the (projective) Hilbert space and each probability measure $\mu_\psi$ with a Dirac $\delta$-function on $\Lambda$ \cite{harrigan2010einstein}.  The ontological models framework also accommodates \emph{hybrid} models  in which the quantum state is both physical and statistical -- indeed the deBroglie-Bohm interpretation fits in this category. Such hybrid interpretations can be sufficient to explain the randomness of experimental outcomes because pure quantum states provide incomplete (statistical) knowledge about the physical variables.  However, the PBR argument is concerned with ontological models in which pure quantum states are not only statistical but \emph{purely statistical}; the necessary condition they adopt for such models is the following \cite{harrigan2010einstein}: there exist at least two distinct non-orthogonal states $\ket{\psi}$ and $\ket{\phi}$ such that $\mu_{\psi}(\lambda)$ and $\mu_{\phi}(\lambda)$ overlap non-trivially (i.e. on a set of non-zero measure). This leads us to a word of caution regarding terminology: the PBR paper uses the term ``statistical" to refer to those models which we call ``purely statistical". Our terminology is chosen to avoid conceptual confusion about the class of interpretations to which the PBR theorem applies.

\textit{PBR argument.}---Suppose there is a preparation device that gives out two quantum states independently, either $\ket{0}$ or $\ket{+}$ such that $|\braket{0|+}|=1/\sqrt{2}$.  

Now suppose that the ontological model underlying quantum mechanics is such that $\mu_{0}(\lambda)$ and $\mu_{+}(\lambda)$ overlap on a set of non-zero measure $\epsilon$. The states $\Ket{0}$ and $\Ket{+}$ span a 2-dimensional Hilbert space $H_0$. Composition of two such systems results in a Hilbert space $H=H_0\otimes H_0$ with $\ket{0}\otimes\ket{0},\ket{0}\otimes\ket{+},\ket{+}\otimes\ket{0}$, and $\ket{+}\otimes\ket{+}$ as a basis. Now introduce an entangled measurement $M$ whose eigenstates $\{\ket{\xi_{00}},\ket{\xi_{0+}},\ket{\xi_{+0}},\ket{\xi_{++}}\}$ form an orthonormal basis for $H$ such that $\braket{\xi_{ab}{|ab}}=0$ for $a,b\in\{0,+\}$. To reach a contradiction with quantum mechanics an additional assumption about the nature of the ontological model is required: 
	\begin{defn} [Preparation Independence]
		\textit{The composition of independently prepared pure states $\ket{\psi}$ and $\ket{\phi}$ results in probability distributions that satisfy}
			\begin{equation}
				\mu_{\psi,\phi}(\lambda_1, \lambda_2)=\mu_{\psi}(\lambda_1)\mu_{\phi}(\lambda_2).
			\end{equation}
	\end{defn} 
\noindent Here $\lambda_i$ is a set of arbitrary ``local'' variables associated with subsystem $i$. The implicit structure associated  with the preparation independence assumption is that a complete account of the composition of the two independent preparations is prescribed by the state space $ \Lambda_\textrm{LHV} = \Lambda_ 1 \times \Lambda_2$.

The first outcome is orthogonal to $\ket{0}\otimes\ket{0}$, thus quantum theory predicts that this outcome has probability zero when the quantum state prepared was $\ket{0}\otimes\ket{0}$. Similarly for the other three states. Assuming the measurement device reads its outcome solely on the ontic states prepared by each system, with probability $\epsilon^{2}$ the measurement device is uncertain which of the four quantum states was prepared. For example, outcome $1$ might occur even if the quantum state was $\ket{0}\otimes\ket{0}$. This contradicts the predictions of quantum mechanics. PBR infer (incorrectly, as we explain below) that this contradiction rules out the possibility of a purely statistical model.

\textit{The problem with PBR.}--- Here we show that the PBR assumption of preparation independence relies on an assumption of ``local causality" which is the likely source of the conflict they obtain with quantum theory. First we remark that the assumption of local causality is widely considered to be the assumption that is ruled out by the conflict between Bell's inequality and quantum theory. In Bell's set-up the systems are initially brought together and prepared in an entangled quantum state, and then separated and subjected to independent measurements.  The key point is that the kind of locality that is ruled out by Bell's inequality is a locality in the modelling of the \emph{independent measurements}! Let $\Pr(k_A,k_B) = \int d\lambda \mu_\psi(\lambda) \xi(k_A,k_B| \lambda, M_A, M_B)$ stand for the joint probability of outcomes $k_A$ and $k_B$ on two space-like separated systems A and B, given independently chosen measurements $M_A$ and $M_B$ on each system. In general $\Pr(k_A,k_B)$ will exhibit correlations. However, if we assume that the observed correlations are entirely a consequence of correlations created at the source, encoded in $\mu_\psi(\lambda)$, which are carried to the measurements via the variables $\lambda$, then we can derive the Bell inequality.  The key step is to demand ``factorizability" of the response function $\xi$.  Specifically, by appealing to ``local causality", we can derive the factorizability condition  $\xi(k_A,k_B| \lambda, M_A, M_B) = \xi(k_A|\lambda, M_A) \xi(k_B| \lambda, M_B)$.  In words, after conditioning on the causal influence due to $\lambda$, we demand that the response function at site $A$ cannot depend on either the choice of measurement $M_B$, nor the outcome $k_B$, for system $B$, which is space-like separated from system $A$, and vice versa.  
That is, the local causality assumption disallows the response functions for the independent measurements to depend on any non-local variables (via the assumptions of outcome independence and parameter independence). To make the parallel even more explicit: the key assumption of Bell's theorem is that the response functions for independent measurements are represented by product  probabilities, or, more precisely, product conditional probabilities. These conditional probabilities take the product form \emph{after} they are conditioned on the \emph{global} variables associated with the entangled preparation of the joint system. If the response functions for the independent measurements are allowed to depend on variables that are not local to the measurement, such as the remote measurement setting or the remote measurement outcome, then the Bell inequality cannot be derived.  In other words, in order to circumvent the Bell inequality, we need only allow the independent measurements to depend on some variables beyond those that are locally accessible. In light of this, we now observe that PBR's assumption of preparation independence is also motivated by an appeal to ``local causality": specifically, PBR assume that independent (space-like separated) preparations must be represented by a probability function that factorizes,  $\mu(\lambda|\psi, \phi) = \mu(\lambda_1|\psi) \mu(\lambda_2|\phi)$. Preparation independence is thus justified by assuming that the outcome $\lambda_1$  can not depend on the choice of preparation $\phi$ for system 2 (parameter independence)  nor should it depend on the (random) outcome $\lambda_2$ for system 2 (outcome independence), and \emph{vice versa}.  These assumptions may seem like weaker ``locality" assumptions than Bell's, because they are applied to independent preparations, which share no ``common cause", rather than  independent measurements, which share a ``common cause" via the source variables $\lambda$ in the Bell scenario. But the key point is that in Bell's scenario, the observed correlations are ``shocking'' precisely because the formalism already accounts for the correlations due to the source via $\mu(\lambda)$; after conditioning on $\lambda$, the response functions are assumed to be fully independent: indeed, this is the very meaning of \emph{conditional independence}. Hence we see that a conflict with quantum predictions arises specifically when independent operations (measurements in the case of Bell and preparations in the case of PBR) are presumed to depend only upon locally accessible variables, and consequently factorize.  In this way PBR's assumption of preparation independence encodes an assumption local causality, but we already know from Bell's theorem that, if we are going to  consider the hidden variable models framework in the first place, then we must reject that assumption.

The above analysis suggests that the conflict between PBR and quantum predictions follows from the way PBR requires us to model independent operations,  rather than the possibility of a purely statistical hidden variable model. Is it possible to identify a weaker notion of independence that is relevant in the broad framework of hidden variable models and consistent with quantum predictions, and, guided by this insight, construct a purely statistical hidden variable model that reproduces the quantum predictions? We achieve both of these aims below.

\textit{Independence in a non-local setting.}---Bell's theorem tells us that any hidden-variable theory of quantum mechanics must violate the principle of ``local causality".  A natural way to achieve this is to allow non-local variables.  Indeed this is precisely how the deBroglie-Bohm interpretation \cite{dbb} reproduces the quantum predictions, and of course also the orthodox interpretation in which the quantum states are presumed to give a  complete account of physical reality \cite{harrigan2010einstein}.   Let's formulate an ontological model that encodes this insight by introducing an additional hidden variable $\lambda_s \in \Lambda_s$  that carries strictly \emph{relational} information about systems 1 and 2. By their very nature these additional, relational variables are presumed to  not be accessible to local measurements -- they are hidden under local measurements.  This non-local hidden variable model (NLHV) has the form $\Lambda_\textrm{NLHV} = \Lambda_{1} \times \Lambda_{2} \times \Lambda_s$,
where the variables $\lambda_i \in \Lambda_{i}$ correspond to the usual local properties particular to  each subsystem, such as the positions and momenta of each particle. The Cartesian product structure assumed above is not necessary for our argument but is a conceptually simple way to introduce a non-local hidden variable. 

Given this more general ontological structure, we still need to accommodate the natural empirical requirement that the local variables $\lambda_i$ associated with independent preparations should reflect the independence  of each local preparation procedure.  We emphasize here that we are considering the variables $(\lambda_s)$ to be inaccessible to local measurements (more generally, to LOCC measurements), and their influence is revealed only under joint (entangling) measurements of systems 1 and 2.  Given that we have a  set of three random variables $(\lambda_1, \lambda_2, \lambda_s)$, there is an important  distinction between full independence and marginal independence that plays a critical role in our model below. In particular, empirical considerations motivate marginal independence, but  are not sufficient to demand full independence. Remarkably, this distinction is enough to overcome the PBR ``no go'' theorem. Consider the following:
\begin{defn}[Local Independence] \textit{The composition of independently prepared pure states $\ket{\psi}$ and $\ket{\phi}$ produce probability distributions that are independent after marginalizing over any inaccessible variables}
\begin{equation}
\int_{\Lambda_{s}}\!\!\!\!\mu_{\psi,\phi}(\lambda_1,\lambda_2,\lambda_{s})\textrm{d}\lambda_{s} = \mu_{\psi}(\lambda_1)\mu_{\phi}(\lambda_2).
\end{equation}
\end{defn}
If we allow for the possibility of an additional hidden variable such as $\lambda_s$ that is not associated with the individual properties of either system, then local independence is the weakest assumption that is consistent with the requirement that it must be possible to give independent descriptions of non-interacting systems.\\
\indent \textit{Explicit NLHV toy model.}---We now give an explicit model that is purely statistical and also satisfies the condition of local independence defined above.\\
\begin{figure}
\centering
\includegraphics[width=80mm,height=40mm]{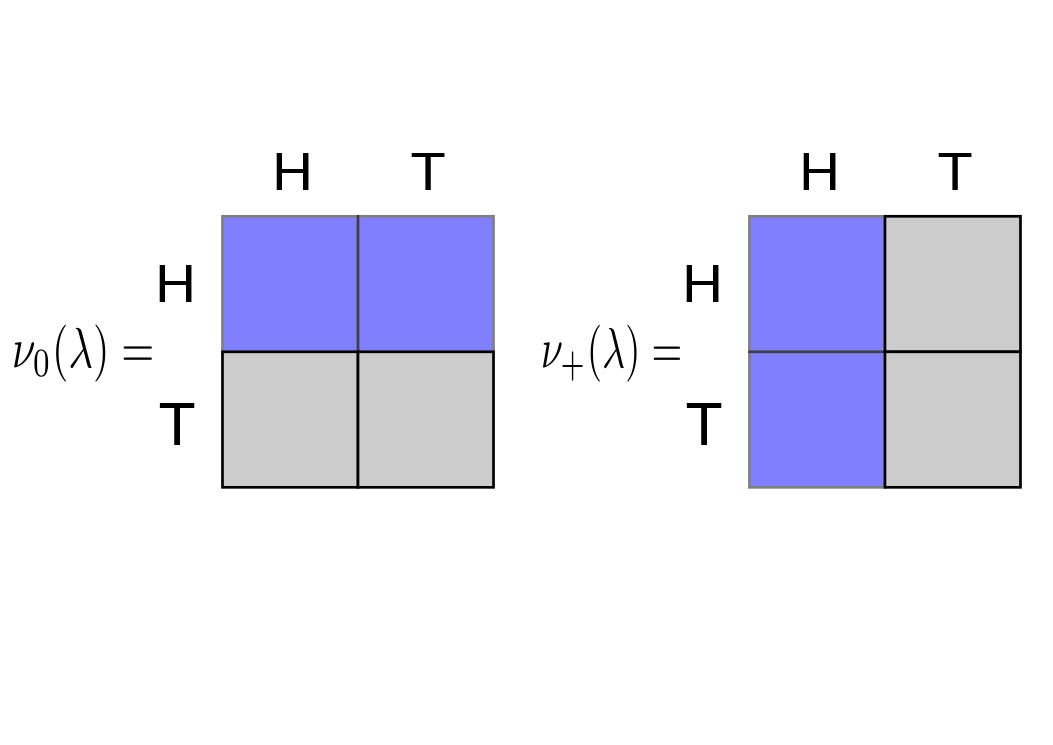}
\caption{The states $\ket{0}$ and $\ket{+}$ are represented by the following probability distributions in our model: when the state $\ket{0}$ is prepared, the underlying ontic state is either $\lambda=(HH)$ or $\lambda=(HT)$, each occurring with probability 1/2. Similarly for the state $\ket{+}$.}
\end{figure}
\indent Consider again the states $\Ket{0}$ and $\Ket{+}$.  For each qubit system, suppose that the ontic space is the space of all possible outcomes of two fair coin flips, i.e. $\Lambda_i=\{H,T\}^2$ for $i \in \{ 1,2\}$. The probability distributions $\nu_0(\lambda)$ and $\nu_+(\lambda)$ corresponding to preparations $\Ket{0}$ and $\Ket{+}$ are given by (cf. Fig. 1):
\begin{gather}
	\nu_{0}(\lambda)=1/2~~\text{if }\ \lambda \in \{\text{(HH), (HT)}\}\\
	\nu_{+}(\lambda)=1/2~~\text{if }\ \lambda \in \{\text{(HH), (TH)}\},
\end{gather}
and zero everywhere else. The composition of two systems introduces an additional non-local hidden variable $\lambda_{s} \in \Lambda_s =\{1,2\}$ so that the composite system is described by a variable $\lambda \in \Lambda$, where
\begin{equation}
	\Lambda =\{H,T\}^2 \times \{H,T\}^2 \times \{1,2\}.
\end{equation}

With this NLHV model, we identify the probability distributions associated with the product states $\Ket{00}$, $\Ket{0+}$, $\Ket{+0}$, and $\Ket{++}$ as (cf. Fig. 2):
\begin{align*}
	&\nu_{00}(\lambda)=\begin{cases}
	1/4 &\text{if } \lambda=\text{(HH,HH,1), (HT,HH,1)}\\
	1/4 &\text{if } \lambda=\text{(HH,HT,1), (HT,HT,1)}
	\end{cases}\\ 
	&\nu_{0+}(\lambda)=\begin{cases}
	1/4 &\text{if } \lambda=\text{(HH,HH,2), (HT,HH,1)}\\
	1/4 &\text{if } \lambda=\text{(HH,TH,1), (HT,TH,1)}
	\end{cases}\\
	&\nu_{+0}(\lambda)=\begin{cases}
	1/4 &\text{if } \lambda=\text{(HH,HH,2), (HH,HT,1)}\\
	1/4 &\text{if } \lambda=\text{(TH,HH,1). (TH,HT,1)}
	\end{cases}\\
	&\nu_{++}(\lambda)=\begin{cases}
	1/4 &\text{if } \lambda=\text{(HH,HH,1), (HH,TH,1)}\\
	1/4 &\text{if } \lambda=\text{(TH,HH,1), (TH,TH,1)}
	\end{cases}.
\end{align*}
The additional bit is a hidden variable that encodes relational information about Alice's and Bob's preparations --  the value of $\lambda_s$  depends on the following: suppose Alice and Bob both obtain local outcome``HH'' (leading to the ontic state ``(HH,HH)" for the composite system), then value of the bit $\lambda_s$  takes on the value ``2'' only if they each obtained ``HH'' via different preparations; in all other cases  $\lambda_s$ has value ``1''.\\
\indent It is easy to see  that our model is a purely statistical model for the states of each subsystem - this follows from the fact that $\nu_0$ and $\nu_+$ intersect non-trivially. Moreover, the NLHV model for the composite system satisfies an even stronger notion of being purely statistical: every pair of non-orthogonal states enjoy a non-trivial overlap.\\
\indent We now show that this purely statistical NLHV can reproduce the quantum predictions.
We construct the following response functions to model the measurement $M$ on the composite space of two qubits:
\begin{align*}
	&\xi_1(\lambda)=\begin{cases}
	\frac{1}{2} &\text{if } \lambda=\text{(HH,HH,2), (HH,TH,1), (TH,HH,1)}\\
	1 &\text{if } \lambda=\text{(TH,TH,1)}
	\end{cases}\\
	&\xi_{2}(\lambda)=\begin{cases}
	\frac{1}{2} &\text{if } \lambda=\text{(HH,HH,1), (HH,HT,1), (TH,HH,1)}\\
	1 &\text{if } \lambda=\text{(TH,HT,1)}
	\end{cases}\\
	&\xi_{3}(\lambda)=\begin{cases}
	\frac{1}{2} &\text{if } \lambda=\text{(HH,HH,1), (HT,HH,1), (HH,TH,1)}\\
	1 &\text{if } \lambda=\text{(HT,TH,1)}
	\end{cases}\\
	&\xi_{4}(\lambda)=\begin{cases}
	\frac{1}{2} &\text{if } \lambda=\text{(HH,HH,2), (HT,HH,1), (HH,HT,1)}\\
	1 &\text{if } \lambda=\text{(HT,HT,1)}
	\end{cases}
\end{align*}
In order for these response functions to satisfy condition (i) for the ontological models framework, we must also impose that $\xi_k(\lambda)=1/4$ for any $\lambda$ outside the union of supports of epistemic states, i.e. outside the space of relevant variables, and zero everywhere else.
\begin{figure}
\centering
\includegraphics[width=80mm,height=40mm]{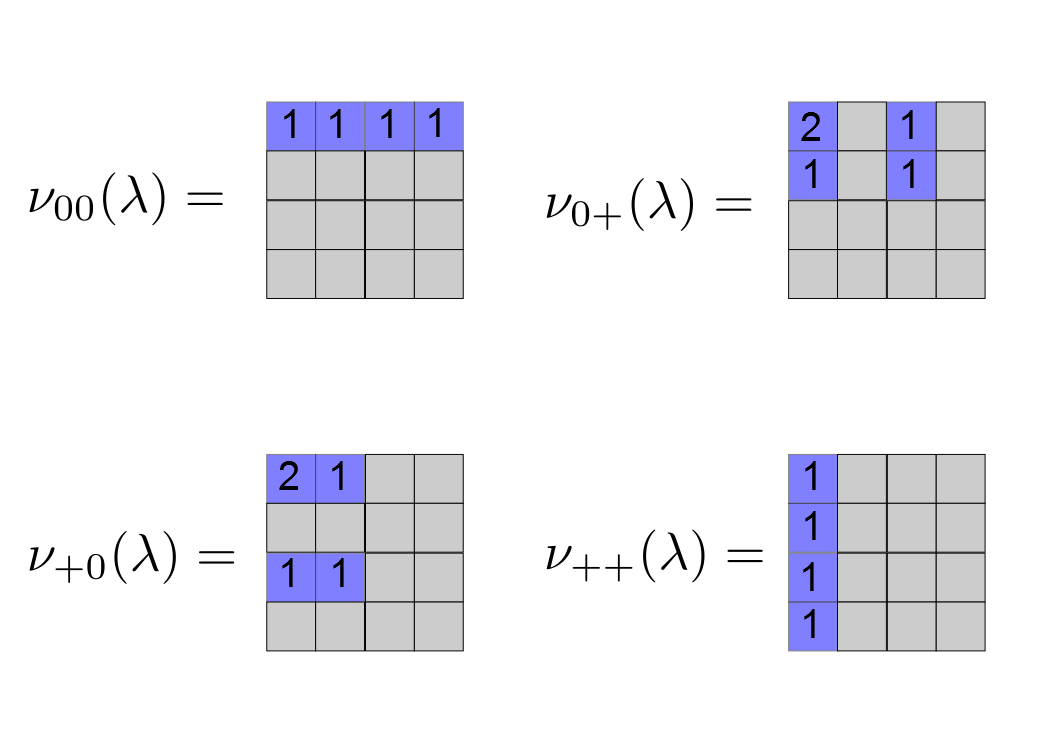}
\caption{The composition of two systems which are either $\ket{0}$ or $\ket{+}$ is represented by the following probability distributions in our ontological model. The numbering on each square is the value of the non-local ontology at that point in the ontic space.}
\end{figure}



\textit{Local Independence vs Preparation Independence.}---We now show that this model satisfies PBR's notion of preparation independence at the level of the local variables of each subsystem. In particular,  we can recover the PBR ``no go'' result by simply denying the possibility that independent operations have support on any variables other than those that are locally accessible.  If we assume that the independent operations are measurements, then this is the assumption of local causality which leads to Bell's inequality. If we assume the independent operations are preparations, then this is the assumption of preparation independence used by PBR.
We can show this explicitly for our toy model by restricting the support of the measurement functions to only the local variables $\lambda_i$, i.e., by assuming that the measurement functions do not depend on the information supplied by the non-local bit. Under this restriction on the support of the measurement functions, we need only consider the marginal probability distributions  on the space $\Lambda_1 \times \Lambda_2$,
which are given by
\begin{align*}
	&\mu_{00}(\lambda_1,\lambda_2)=\begin{cases}
	\frac{1}{4}  &\text{if } (\lambda_1,\lambda_2)=\text{(HH,HH), (HT,HH)}\\
	\frac{1}{4}  &\text{if } (\lambda_1,\lambda_2)=\text{(HH,HT), (HT,HT)}
	\end{cases}\\ 
	&\mu_{0+}(\lambda_1,\lambda_2)=\begin{cases}
	\frac{1}{4}  &\text{if } (\lambda_1,\lambda_2)=\text{(HH,HH), (HT,HH)}\\
	\frac{1}{4}  &\text{if } (\lambda_1,\lambda_2)=\text{(HH,TH), (HT,TH)}
	\end{cases}\\
	&\mu_{+0}(\lambda_1,\lambda_2)=\begin{cases}
	\frac{1}{4}  &\text{if } (\lambda_1,\lambda_2)=\text{(HH,HH), (HH,HT)}\\
	\frac{1}{4}  &\text{if } (\lambda_1,\lambda_2)=\text{(TH,HH), (TH,HT)}
	\end{cases}\\
	&\mu_{++}(\lambda_1,\lambda_2)=\begin{cases}
	\frac{1}{4}  &\text{if } (\lambda_1,\lambda_2)=\text{(HH,HH), (HH,TH)}\\
	\frac{1}{4}  &\text{if } (\lambda_1,\lambda_2)=\text{(TH,HH), (TH,TH)}
	\end{cases}
\end{align*}
and zero everywhere else. It is easy to check that the marginal probability distributions satisfy preparation independence. 
Under this assumption of local causality, the PBR contradiction follows directly: there exists some $\epsilon>0$ such that the ontic state  prepared by one of the systems comes from the overlap region $\Delta\subset \Lambda$ of the probability distributions $\nu_0(\lambda)$ and $\nu_+(\lambda)$. So at least $\epsilon^2$ of the time, the measurement device is unsure which of the quantum states $\ket{00}$, $\ket{0+}$, $\ket{+0}$, or $\ket{++}$ was prepared. Therefore the quantum statistics will \textit{not} be reproduced under these assumptions. 

\textit{Discussion.}---Non-locality is a feature of hidden variable models of quantum mechanics  that is known to be sufficient to account for experimental scenarios in which the quantum state is entangled and the measurements are chosen independently.  Our analysis of the PBR argument shows that a purely statistical model with non-local hidden variables can also account for the quantum predictions in experimental scenarios in which the quantum states are chosen independently and the quantum measurement is entangled. In this context we have proposed  ``local independence'', which is a weaker form of independence consistent with the requirement that non-interacting spatially separated systems admit independent descriptions. Interestingly, our notion of \emph{local independence} is empirically indistinguishable from preparation independence under local measurements within our model, because we assume that the non-local variables are inaccessible to such measurements.  It is an exciting open problem to determine whether an independence requirement of this type, within the framework of non-local hidden variable models, can explain the violation of local causality implied by Bell's theorem. 

Whereas the original PBR paper and subsequent papers \cite{Nigg2012,Barrett2013} have begun characterizing the PBR result as an implication that purely statistical interpretations are untenable, our work suggests that this inference is invalid.  On the contrary, PBR's preparation independence assumption corresponds to an assumption of local causality (applied at the level of independent preparations rather than independent measurements). But we have already known, for 49 years \cite{Bell64}, that this assumption is violated by hidden variable models of quantum mechanics.  By Occam's razor it would appear that  the assumption of a  purely statistical interpretation is actually an innocent bystander in the PBR argument, rather than the driving force behind their contradiction. 

\textit{Note.}---After completion of this work we became aware of Ref.~\cite{SF2013} which makes a different but related criticism of PBR's preparation independence assumption.


\textit{Acknowledgements.}--- We thank Joel Wallman,  Terry Rudolph, Matt Leifer,  Matt Pusey, Steve Bartlett, Hillary Carteret  and Darren Wilson for helpful comments.  
\vspace{-0.2in}
\bibliography{PBRPaper}

\end{document}